\documentclass[%
 reprint,
 superscriptaddress,
 amsmath,amssymb,
 aps,prapplied
]{revtex4-2}
\usepackage{graphicx}
\usepackage{dcolumn}
\usepackage{bm}
\usepackage{multirow}
\usepackage[T1]{fontenc}
\usepackage{graphicx}  
\usepackage{amsmath}  
\usepackage{comment}
\usepackage{xcolor} %
\usepackage{amsfonts} 
\begin{document}
\title{Complete characterization of beam deflection based on double weak value amplification system }
\author{Yu Wang}
\affiliation{State Key Laboratory of Quantum Optics Technologies and Devices, Taiyuan 030006, China}
\affiliation{College of Physics and Electronic Engineering, Shanxi University, Taiyuan 030006, China}
\author{Rongguo Yang}
\email{yrg@sxu.edu.cn}
\affiliation{State Key Laboratory of Quantum Optics Technologies and Devices, Taiyuan 030006, China}
\affiliation{College of Physics and Electronic Engineering, Shanxi University, Taiyuan 030006, China}
\affiliation{Collaborative Innovation Center of Extreme Optics, Shanxi University, Taiyuan, 030006, China}
\author{Jing Zhang}
\affiliation{State Key Laboratory of Quantum Optics Technologies and Devices, Taiyuan  030006, China}
\affiliation{College of Physics and Electronic Engineering, Shanxi University, Taiyuan 030006, China}
\affiliation{Collaborative Innovation Center of Extreme Optics, Shanxi University, Taiyuan, 030006, China}
\author{Chenzhen Luo}
\affiliation{State Key Laboratory of Quantum Optics Technologies and Devices, Taiyuan 030006, China}
\affiliation{College of Physics and Electronic Engineering, Shanxi University, Taiyuan 030006, China}
\author{Xiaomin Liu}
\affiliation{State Key Laboratory of Quantum Optics Technologies and Devices, Taiyuan 030006, China}
\affiliation{College of Physics and Electronic Engineering, Shanxi University, Taiyuan 030006, China}
\author{Kui Liu}
\affiliation{State Key Laboratory of Quantum Optics Technologies and Devices, Taiyuan 030006, China}
\affiliation{Collaborative Innovation Center of Extreme Optics, Shanxi University, Taiyuan, 030006, China}
\author{Jiangrui Gao}
\affiliation{State Key Laboratory of Quantum Optics Technologies and Devices, Taiyuan 030006, China}
\affiliation{Collaborative Innovation Center of Extreme Optics, Shanxi University, Taiyuan, 030006, China}
\begin{abstract}
The precise measurement of spatial attitude parameters is critical for applications in inertial navigation, industrial monitoring, instrument calibration, quantum metrology, etc. In this work, we theoretically investigate and experimentally realize the simultaneous measurement of the yaw and pitch angles using a Hermite-Gaussian-postselected double weak value system integrated with two sets of high-order-mode balanced homodyne detections, thereby achieving a complete characterization of the beam deflection. Signals of the yaw and pitch angles that are involved in TEM$_{10}$ and TEM$_{01}$ modes output from two dark ports of the system can be measured independently. As a result, the obtained minimum measurable yaw and pitch angles of beam deflection are 83 prad and 89 prad, respectively. This work expands the beam deflection measurement to two dimensions, providing new insights for future high-precision multi-parameter spatial precise detection.
\end{abstract}
\maketitle
\section{Introduction}
High-precision measurement of spatial parameters, as a cornerstone technology in optical metrology \cite{meyer1988novel,yang2016far}, industrial monitoring \cite{yang1997measurement}, and quantum sensing \cite{taylor2013biological,pooser2015ultrasensitive,delabert2008quantum}, has been widely applied in tracking of a vibrating structure \cite{huang2023spatial}, micro-nano structure deformation detection \cite{vella2019simultaneous}, and precision instrument calibration \cite{gang2010application}. Increasing the number of incident photons is a conventional way to improve measurement accuracy but is limited by detector saturation \cite{giovannetti2004quantum}. The weak value amplification (WVA) technique not only circumvents detector saturation \cite{xu2020approaching,harris2017weak} but also suppresses detector jitter noise \cite{jordan2014technical,viza2015experimentally,sinclair2017weak,harris2017weak,feizpour2011amplifying}, by measuring few photons at the dark port of the interferometer \cite{jordan2014technical}. Since the WVA technique can achieve higher accuracy than the direct detection, it has been widely applied in the precision measurement of many parameters, such as beam deflection \cite{dixon2009ultrasensitive}, displacement \cite{xu2020approaching}, angular rotation \cite{magana2014amplification}, and longitudinal phase \cite{brunner2010measuring}. Various methods based on the WVA technique have been explored to improve measurement accuracy \cite{lyons2015power,wang2016experimental,zhang2023small,zhang2024precision}. The power recycling technique improved the signal-to-noise ratio (SNR) in weak-value-based beam-deflection measurement \cite{lyons2015power,wang2016experimental}. Combining the WVA technique and the balanced homodyne detection (BHD) strengthened each other’s advantages and performed better than the split detection \cite{zhang2023small}. Quantum precision measurement beyond the shot-noise limit was realized using the squeezing-assisted WVA technique in the optical small-tilt measurement \cite{zhang2024precision}. The above experiments based on the WVA technique are mainly focused on single-parameter estimation, while the simultaneous estimation of multiple parameters is also indispensable in quantum metrology, quantum imaging, and biological sensing, etc. An extension of the WVA technique to simultaneous measurement of multiple parameters, critical dimension and orientation of the sub-wavelength structured sample, was presented with polarization-controlled pre-selection and post-selection \cite{vella2019simultaneous}. For estimating the spatial displacement and angular tilt of light at the same time, a scheme based on high-order Hermite-Gaussian (HG) states was proposed, achieving experimental precision of 1.45 nm and 4.08 nrad, respectively \cite{xia2023toward}. Note that only one-dimensional situations (yaw angles) are considered in all the above-mentioned beam deflections; however, complete characterization of both yaw and pitch angles is also required in many actual application scenarios. In this work, a double weak value amplification precision measurement system with post-selection of spatial high-order modes is proposed, which enables the simultaneous measurement of the yaw and pitch angles–corresponding to the two-dimensional beam deflection.
\section{EXPERIMENTAL SETUP AND THEORETICAL DERIVATION}
The experimental setup is shown in Fig.1. A TEM$_{00}$ mode beam from the laser source is divided into two parts by a polarized beam splitter (PBS). One part passes through a mode-converter and becomes a TEM$_{10}$ mode beam, which provides the local beams for two sets of high-order-mode BHD systems. The other part passes through a mode-cleaner and becomes a purer TEM$_{00}$ mode with suppressed extra noise, then enters a double weak value amplification system (including a Sagnac interferometer and an unbalanced Mach-Zehnder (UMZ) interferometer) as the pointer state.
\begin{figure}
    \centering
\includegraphics[width=1\linewidth]{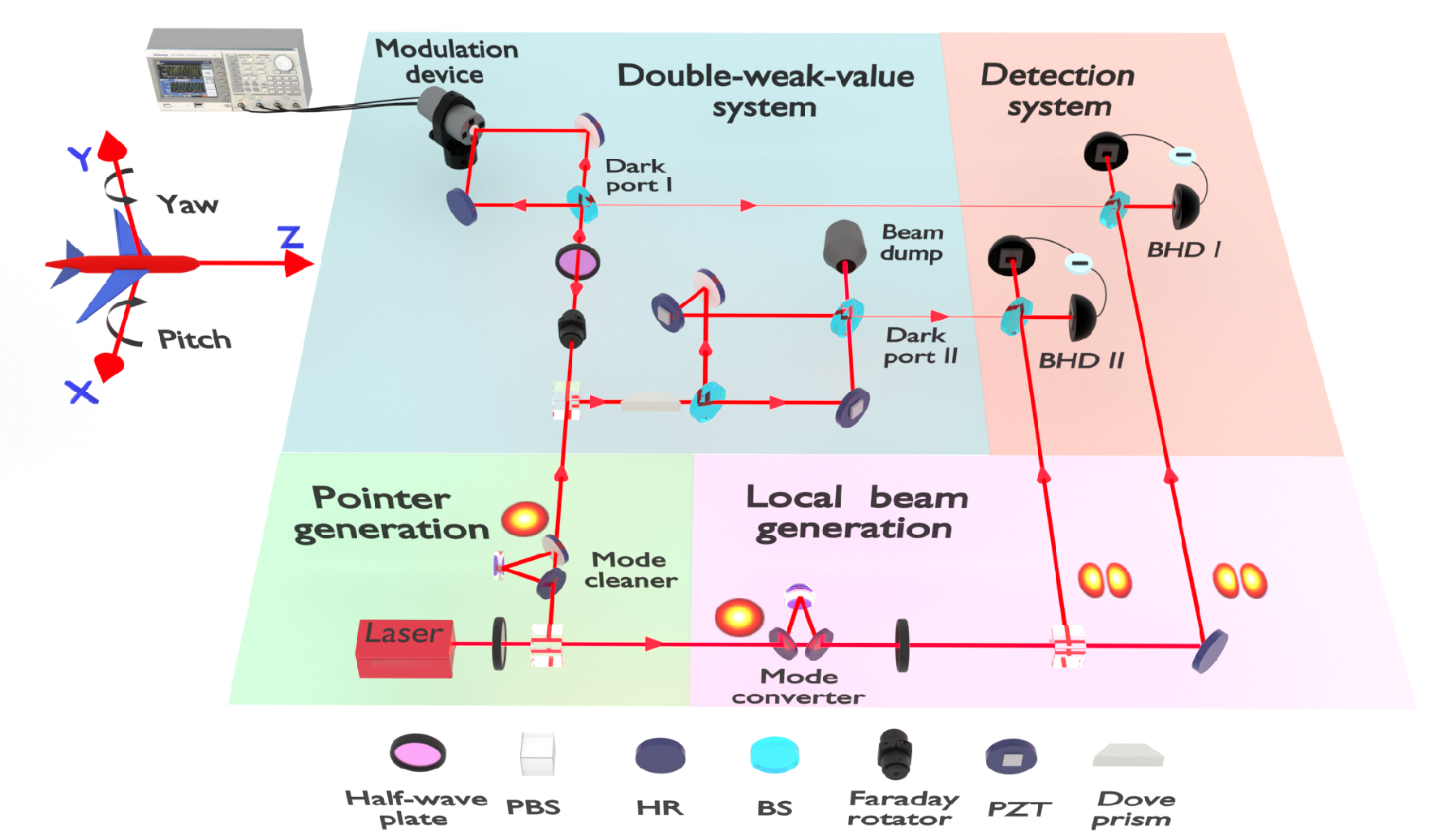}
    \caption{The experimental configuration for simultaneously measuring  yaw and pitch angles using a double weak value amplification system incorporating two high-order-mode BHD systems.}
\end{figure}
The pointer state goes through the Faraday rotator and enters the Sagnac interferometer, where a mirror is fixed on an s-330 (PI, Germany) piezoelectric actuator and induces a beam deflection due to the modulation of the driving signal. Note that the s-330  actuator consists of four piezoelectric transducers (PZT) distributed at the vertex positions of a square structure, and the yaw angle $\theta$ and the pitch angle $\varphi$ can be manipulated by altering the frequencies of the horizontal and vertical PZT pairs. The corresponding displacements can be obtained by certain geometric relations:  $d_{\theta}=(l/2)\sin\theta$ and $d_{\varphi}=(l/2)\sin\varphi$, where $l=19mm$ is the distance between the horizontal and vertical PZT pairs. The output beam from the dark port I carries the information of the yaw angle and can be measured by the BHD I system. The other output beam of the Sagnac interferometer backs along the original path, passes through the Dove prism, then enters the UMZ interferometer. The output beam from the dark port II of the UMZ interferometer carries the pitch angle information and is measured by the BHD II system. According to the above experimental process, the corresponding theoretical analysis can be derived. The pointer state is a TEM$_{00}$ mode beam propagating along z-axis, whose transverse spatial distribution ${\psi _0}\left( {x,y} \right)$ can be expressed as
\begin{align}
\psi_{0}(x,y) = \left( \frac{2}{\pi \omega_{0}^2} \right)^{\!\!\frac{1}{2}}
& H_{0}\left( \frac{x}{\sqrt{2}\,\omega_{0}} \right) 
H_{0}\left( \frac{y}{\sqrt{2}\,\omega_{0}}\right) 
e^{ -\frac{{x+y}^{2}}{4\omega_{0}^{2}}},
\end{align}
where $\omega_x^2$ and $\omega_y^2$ represent the variance of spatial distribution of fundamental Gaussian state in the horizontal and vertical directions, which can be replaced by $\omega_{0}^2$ due to the beam ellipticity of over 99\% in our experiment. $H_0$ is the 0th-order Hermite polynomial. The yaw and pitch motions of the mirror fixed on the S-330 actuator, are equivalent to the momentum modulation of the beam in x and y dimensions, which can be described as: $k_x=2\pi\sin \theta/\lambda$ and $k_y = 2\pi\sin \varphi/\lambda$. Then the deflected TEM$_{00}$ beam $\psi_0^{\theta,\varphi}(x,y)$ can be expressed as:
\begin{align}   \psi_0^{\theta,\varphi}(x,y) &=\exp(ik_x x)\exp(ik_y y)\psi_0(x,y)\nonumber\\ &\approx (1 + i k_x x + i k_y y)\psi_0(x,y). 
\end{align}
Considering the relationship between the 0th and 1st order modes, i.e., $\psi_0(x) = \omega_0 \psi_1(x)/x$, $\psi_0(y) = \omega_0 \psi_1(y)/y$, it can be rewritten as:
\begin{equation}
\psi_0^{\theta,\varphi}(x,y) \approx \psi_0(x,y) + i \omega_0 k_x \psi_1(x) + \\i \omega_0 k_y \psi_1(y),
\end{equation}
where $\psi_1(x)$ and $\psi_1(y)$ represent the first-order modes in the horizontal and vertical directions, respectively. Generally, a standard weak measurement process consists of three steps: pre-selection, weak interaction, and post-selection. Given that $x$- and $y$- directions are orthogonal and independent, the measurements of the pitch and yaw angle can be implemented separately. On the one hand, the signal of the yaw angle $\theta$  can be obtained when constructive interference of the induced TEM$_{10}$ mode occurs at the dark port I. There are two possible paths in the Sagnac interferometer: the clockwise path $\left|\circlearrowright\right\rangle$  and the counterclockwise path $\left|\circlearrowleft\right\rangle$. After completing a full cycle, the two beams return to the input beam splitter (BS), exhibiting a relative phase difference $\phi_1$ between them the two paths. Assume $\lvert i \rangle$ is the pre-selection state of the system, it can be expressed by $\lvert i \rangle =\left( e^{-i \phi_1/2} \lvert \circlearrowright  \rangle + e^{i \phi_1/2} \lvert \circlearrowleft \rangle \right)/\sqrt{2}$. The Hamiltonian of the weak interaction in the horizontal direction can be expressed as: $H_{\text{int}}^\theta = g_\theta\hat{A}_\theta k_x \hat{x} $, where ${g_\theta }$ is the coupling coefficient, and ${\hat A_\theta }$ is the corresponding measurement operator on the system. By altering the relative phase $\phi_1$, constructive and destructive interference of the TEM$_{00}$ and  TEM$_{01}$ modes can be achieved at the bright and dark output ports, respectively. The pointer state $|\psi_0(x)\rangle$ and  pre-selection state $| i \rangle$  couple as the direct product state $|i \rangle |\psi_0(x)\rangle$. The system evolves under the unitary evolution operator $\hat{U}{_\theta}=e^{ - i\int{{ H_{\text{int}}^\theta}dt}}$, so the output  state at the dark port I can be expressed as
\begin{align}
|\psi_{ f_-}^\theta\rangle
    &= \langle f_-|{{\hat U}_\theta }\left| i \right\rangle \left| {{\psi _0}(x)} \right\rangle \nonumber \\[1ex]
    &\approx \langle f_- |i\rangle \left( 1 - iA_{\rm{w}}^\theta {k_{\rm{x}}}\hat{x} \right)\left|{{\psi _0}(x)} \right\rangle ,
\end{align}
\begin{figure}
    \centering
    \includegraphics[width=1\linewidth]{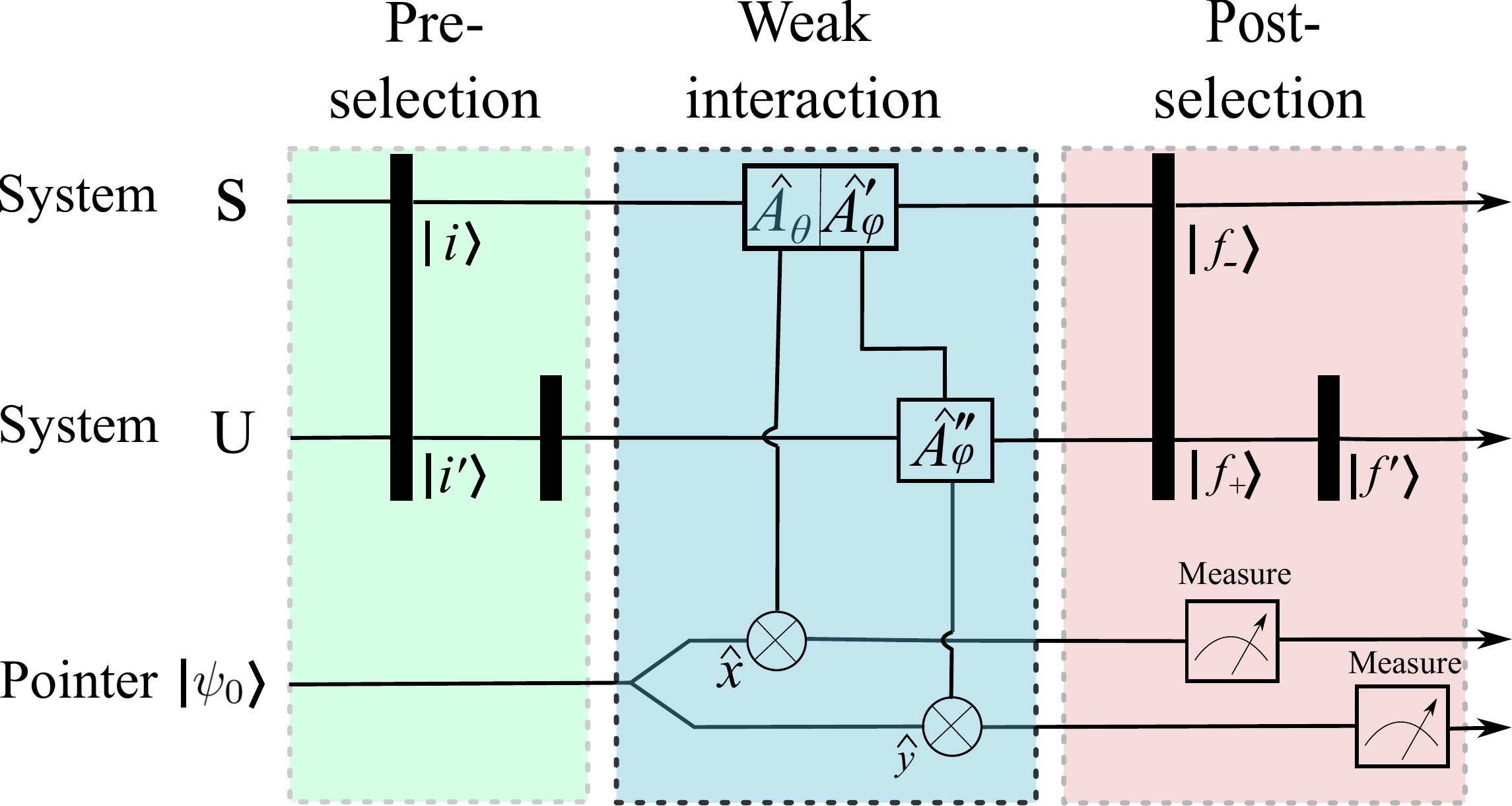}
    \caption{Schematic diagram of the double weak value amplification system based on the HG post-selection. \textbf{S} and \textbf{U} denote the Sagnac and UMZ interferometers, respectively.}
\end{figure}
where the weak value $A_\mathrm{w}^\theta$ is defined by: $A_\mathrm{w}^\theta = \frac{\langle f_- \lvert\hat{ A}_\theta \rvert i \rangle}{\langle f_- \lvert i \rangle}$, the measurement operator is ${\hat A_\theta } = (\left| \circlearrowright \right\rangle \left\langle \circlearrowright  \right| - \left|\circlearrowleft    \right\rangle \left\langle  \circlearrowleft\right|)$, and the post-selection state is $\left|f_\pm\right\rangle =(\left| \circlearrowright \right\rangle  \pm \left|\circlearrowleft    \right\rangle )/\sqrt{2}$, where "$+$" and "$-$" correspond to the bright and dark ports, respectively. Then the post-selection probability ${p}_\theta$ at the dark port I is ${p}_\theta=\left| {\left\langle f_- \right|\left. i \right\rangle } \right|^2=\sin^2({\phi_1/2 })= {P}_{out}^\theta/{{P}_{in}}$. Note that the first and second terms in the second row of the Eq. (4) correspond to the TEM$_{00}$ and TEM$_{10}$ modes, respectively. On the other hand, the pitch angle $\varphi$ in the vertical direction is induced at the same time and same position with the yaw angle $\theta$. At the dark port of  the UMZ interferometer, destructive and constructive interference occur for the TEM$_{00}$ and TEM$_{10}$ modes, respectively. Note that the Dove prism transfers TEM$_{01}$ mode into TEM$_{10}$ mode. There are also two possible paths in the UMZ interferometer: the transmissive path $\lvert T \rangle$ and the reflective path $\lvert R \rangle$. The relative phase difference $\phi_2$ is locked by PZTs fixed on mirrors. It is worth mentioned that the  pre- and post-selection states of the WVA process for measuring pitch angle $\varphi$ is not only depend on the UMZ interferometers, but also depend on the previous Sagnac interferometer. The pointer state $|\psi_0(y)\rangle$ and  pre-selection state $\lvert i \rangle \lvert i' \rangle$ couple as the direct product state $|i \rangle|i'\rangle \psi_0(y)$,  with $\lvert i' \rangle=\left( e^{-i \phi_2/2} \lvert T  \rangle + e^{i \phi_2/2} \lvert R \rangle \right)/\sqrt{2}$ . The Hamiltonian of the weak interaction in the vertical direction is $ H^{\varphi}_{\text{int}} = g_{\varphi} \hat{A}_\varphi k_y \hat{y}$, where $g_{\varphi}$ is the coupling coefficient,  and $\hat{A}_\varphi = \hat{A}''_\varphi\hat{A}'_\varphi$ is the corresponding measurement operator on the system, with $\hat{A}'_\varphi = (\left| \circlearrowright \right\rangle \left\langle \circlearrowright  \right| + \left|\circlearrowleft    \right\rangle \left\langle  \circlearrowleft\right|)$ and $\hat{A}''_\varphi  = (\left|T\right\rangle \left\langle T  \right| - \left|R  \right\rangle \left\langle R\right|)$. In the same way, the post-selection state depend on both  
$\lvert f_+ \rangle =  \left( \lvert \circlearrowright \rangle + \lvert \circlearrowleft \rangle \right)/\sqrt{2}
$ (the bright port of the Sagnac interferometer) and $
\lvert f' \rangle = \left( \lvert T\rangle - \lvert R \rangle \right)/\sqrt{2}$ (the dark port of the UMZ interferometer). Then, the output  state at the dark port II can be expressed as:
\begin{align}
|\psi_{f_+f'}^\varphi\rangle
&=\langle f'|\langle f_+ | \hat{U}_\varphi \big| i \big\rangle \big| i' \big\rangle \big| \psi_0(y) \big\rangle \nonumber\\
&\approx\langle f'|  \langle f_+ | i\rangle|i' \rangle  \left( 1 - i A_w^\varphi k_y \hat{y} \right)|\psi_0(y) \rangle ,
\end{align}
where ${\hat{U}_\varphi}=e^{ - i\int{{ H_{\text{int}}^\varphi}dt}}$ and ${A}_\mathrm{w}^{\varphi} = \frac{\langle f'|\langle f_+\lvert \hat{A}_\varphi\lvert i \rangle\lvert i'\rangle}{\langle f'|\langle f_+\lvert i \rangle\lvert i' \rangle}$.  The post-selection probability $p_{\varphi}$ is   $p_{\varphi}=P_{out}^\varphi/( P_{in}\eta)$, where $\eta$=90\% is the light transmission efficiency between the two interferometers. Substituting Eq. (4) and (5) into Eq. (2) , the final states of the double weak value amplification system can be obtained as  
\begin{subequations}
\begin{align}
|\psi_{f_-}^\theta\rangle &= |\psi_0(x)\rangle - i\omega_0 A_{\mathrm{w}}^\theta k_x | \psi_1(x)\rangle  / \sqrt{p_\theta},\\
|\psi{_{f_+f'}^\varphi}\rangle &= |\psi_0(y)\rangle - i\omega_0 A_{\mathrm{w}}^{\varphi} k_y | \psi_1(y)\rangle/ \sqrt{p_\varphi}.
\end{align}
\end{subequations}
The whole double weak value amplification measurement scheme is shown in Fig.2. The photon number difference between two detectors can be obtained as \cite {delaubert2006tem,sun2014small}
\begin{subequations}
\begin{align}
{{\hat{N}}_\theta^-} &= \sqrt{N_{LO}} \left( 2\sqrt{N_\theta} A_{\mathrm{w}}^\theta k_x \omega_0 + \delta\hat{X}_s^-\right),\\
{{\hat{N}}_\varphi^-} &= \sqrt{N_{LO}} \left( 2\sqrt{N_\varphi} A_{\mathrm{w}}^\varphi k_y \omega_0 + \delta\hat{X}_s^- \right).
\end{align}
\end{subequations}
${N_{LO}}$ denotes the photon number of the local beam, $N_{\theta}$ and $N_{\varphi}$ denote the photon number at the two dark ports, respectively. The quantum noise is $\delta \hat{X}_s^- =(\delta a - \delta a^\dagger)/i$, where ${\hat a^\dagger }$ and $\hat a$ are the creation and annihilation operators, respectively. Note that the first and second terms of the equations correspond to the signal and the noise parts, respectively. Since  $\delta\hat{X}_s^-=1$ for coherent light, the signal-to-noise ratio (SNR) of the measurement is 
\begin{subequations}
\begin{align}
\text{SNR}^{\theta} &= \left(2\sqrt{N_{in}p_\theta}{A_{\mathrm{w}}^\theta} k_x \omega_0 \right)^2= \left( 2\sqrt{N_{\theta}}{A_{\mathrm{w}}^\theta} k_x \omega_0 \right)^2,\\
\text{SNR}^{\varphi} &= \left(2\sqrt{N_{in}\eta  p_\varphi} {A_{\mathrm{w}}^{\varphi}} k_y \omega_0 \right)^2 = \left(2\sqrt{N_{\varphi}} {A_{\mathrm{w}}^{\varphi}} k_y \omega_0 \right)^2.
\end{align}
\end{subequations}
 $A_\mathrm{w}^\theta$ and $A_\mathrm{w}^\varphi$ are the amplification factors of the pitch and yaw angles (signals), respectively. Actually, during the WVA process, the weight of the 1st-order-mode photon numbers that contain the signal information is amplified. Taking SNR=1, the minimum measurable yaw and pitch angles can be obtained,  
\begin{subequations}
\begin{align}
\theta _{\min } &= \frac{\lambda }{4\pi \omega_0 \sqrt{N_{\text{in}}} \cos (\phi_1/2)},\\
\varphi _{\min }&= \frac{\lambda }{4\pi \omega_0 \sqrt{N_{\text{in}} \eta}\cos (\phi_2/2)},
\end{align}
\end{subequations}
where for $\varphi _{\min }$, $\phi_1$ is already included in $\eta$.
\section{EXPERIMENTAL RESULTS}
\begin{figure}
    \includegraphics[width=1\linewidth]{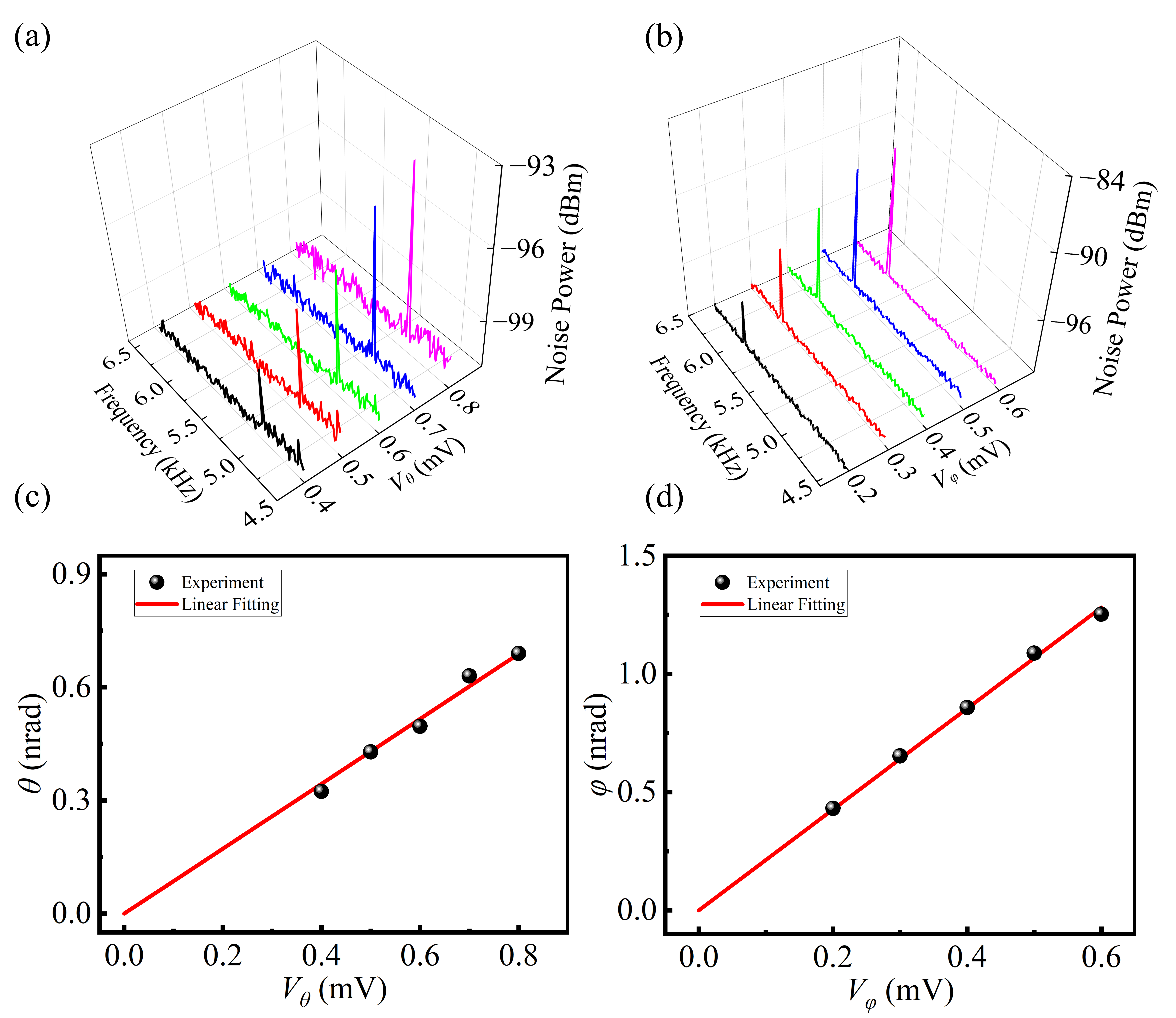}
    \caption{Experimental results for an input power of 50 µW. The noise power spectra obtained from dark ports I and II under different PZT driving voltages are shown in (a) and (b). The linear relationships between the yaw and pitch angles and the PZT driving voltage are depicted in (c) and (d). The local beam power is 1 mW, and the resolution bandwidth (RBW) is 10 Hz.}
\end{figure}
\begin{figure}
    \includegraphics[width=1\linewidth]{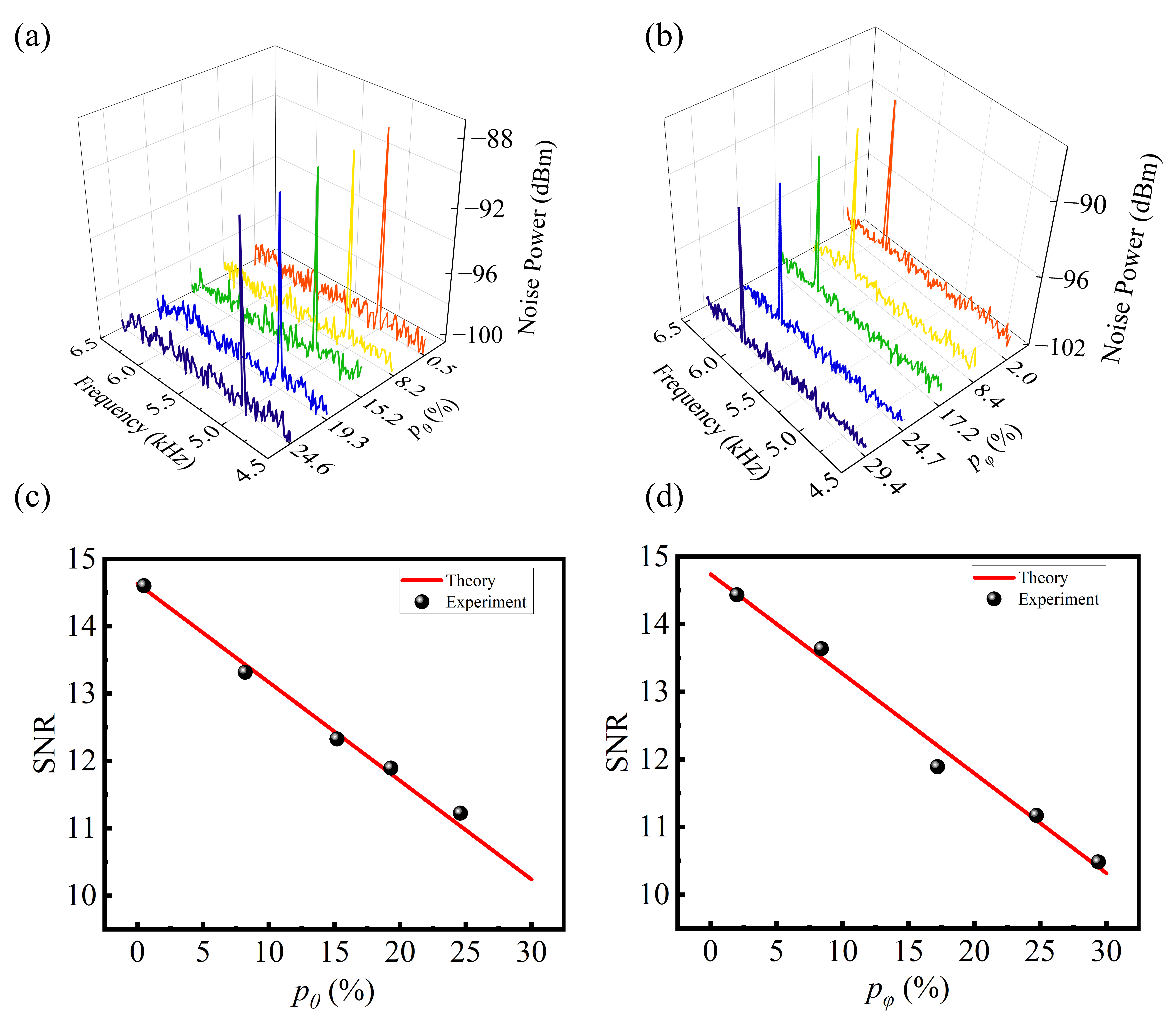}
    \caption{Experimental results for different post-selection probabilities. The noise power spectra obtained from dark port I and II under different post-selection probabilities are shown in (a) and (b), respectively. The corresponding  of signal peak SNRs and theoretical predictions are shown in (c) and (d). The output powers from dark port I and II are set as (0.25 µW (0.5\%), 4.12 µW (8.2\%), 7.63 µW (15.2\%), 9.65 µW (19.3\%), 12.3 µW (24.6\%)) and (0.91 µW (2.0\%), 3.78 µW (8.4\%), 7.76 µW (17.2\%), 11.22 µW (24.7\%), 13.25 µW (29.4\%)), respectively.}
\end{figure}
In the experiment, the yaw and pitch signals are detected under a certain input power, then the system is optimized to obtain higher measurement precision. The PZT pairs responsible for generating the yaw and pitch signals are driven at 5kHz and 6kHz, respectively. 

For an input power of 50 µW, the noise power spectra from two dark ports under different PZT driving voltages are shown in Fig.3(a) and Fig.3(b), with post-selection probability $ p_\theta $ and $p_\varphi$ being 13\% and 20\%, respectively. According to the peak values in Fig.3(a) and Fig.3(b), the measured yaw and pitch angles as functions of the PZT driving voltage are presented in Fig.3(c) and Fig.3(d), respectively. Clearly, the yaw and pitch angle signals are proportional to the PZT driving voltage and can be measured simultaneously. According to the black curves in Fig.3(a) and Fig.3(b) (3dB SNR), the minimum measurable yaw and pitch angles are determined to be 0.32 nrad and 0.43 nrad, induced by 0.4 mV and 0.2 mV driving voltages, respectively. 

Consequently, with the input power fixed at 50 µW, the output power at the dark port is reduced. The noise power spectra detected by two BHD systems under different output powers (corresponding to different post-selection probabilities) are shown in Fig.4(a) and Fig.4(b). The corresponding signal peak SNRs for the yaw and pitch angles as functions of post-selection probabilities are presented in Fig.4(c) and Fig.4(d), respectively. Notably, the SNR is inversely proportional to the post-selection probability, in agreement with Eq. (8). The minimum achievable post-selection probabilities at dark ports I and II are 0.5\% and 2\%, respectively.

\begin{figure}
    \includegraphics[width=1\linewidth]{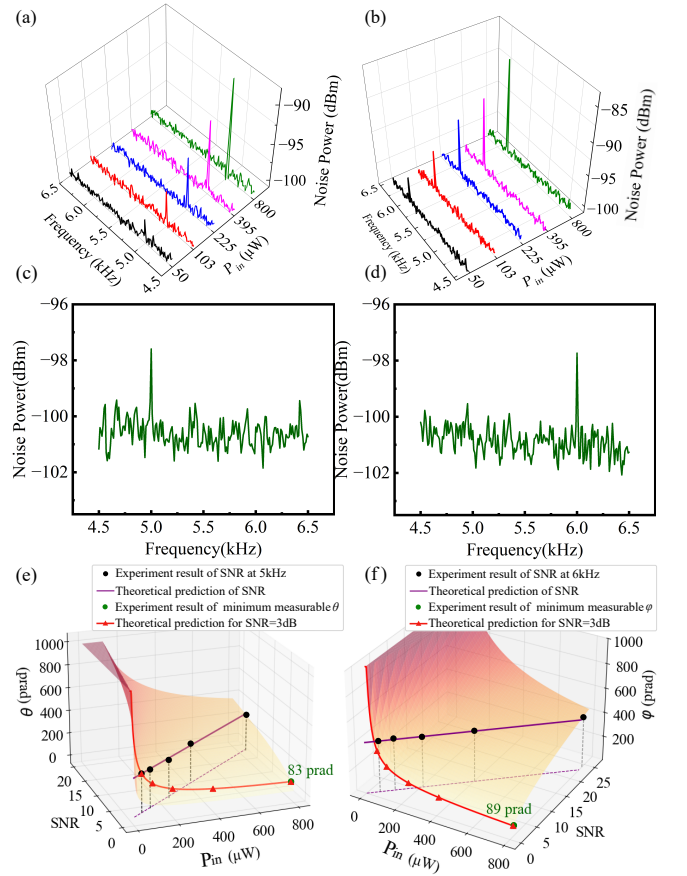}
    \caption{Experimental results and theoretical predictions for measuring yaw ((a), (c), (e)) and pitch ((b), (d), (f)) angles. The noise power spectra under different input powers are shown in (a) and (b). The noise power spectra at 3 dB SNR for an input power of 800 µW are presented in (c) and (d). The integrated experimental and theoretical results are summarized in (e) and (f).}
\end{figure}

Finally, keeping the driving voltage and the minimum post-selection probability, the noise power spectra with different input power, detected by two sets of BHD systems, are shown in Fig.5(a) and Fig.5(b). As the input power increases from 50 µW  to 800 µW, the signal peak values for the yaw and pitch angles rise from 2.96 dB and 3.87 dB to 12.13 dB and 14.10 dB, respectively. Next, to achieve a 3 dB SNR, the voltages applied to the PZT pairs are reduced to 0.1 mV and 0.04 mV, and the resulting noise power spectra are shown in Fig.5(c) and Fig.5(d), from which the minimum measurable yaw and pitch angles are determined to be 83 prad and 89 prad, respectively. The experimental and theoretical results of measuring yaw and pitch angles are comprehensively presented in Fig.5(e) and Fig.5(f). The black and green dots represent the experimental data for the signal peak SNR and the minimum measurable yaw/pitch angles, respectively. The purple lines and the red curves (red triangles denote the data points at different input powers) are the corresponding theoretical predictions. Clearly, the SNR and precision can be enhanced by increasing the input power, and the best minimum measurable yaw and pitch angles are 83 prad and 89 prad, respectively. The corresponding displacements are 0.79 pm and 0.85 pm. 
\section{CONCLUSION }
In this paper, we present a complete experimental characterization of two-dimensional beam deflection, as well as a theoretical analysis using a double weak value amplification system with Hermite-Gaussian post-selection. The detected beams output from two different dark ports are detected by two sets of independent BHD systems, achieving minimum measurable deflections of 83 prad (yaw) and 89 prad (pitch), respectively. This work demonstrates a complete characterization of beam deflection based on a double weak value amplification system, an approach that can be expanded to multi-parameter measurements. For future work, we plan to employ higher-order or squeezed injected beams to further improve measurement precision. 
\section{ACKNOWLEDGMENTS}
This work was supported by the National Natural Science Foundation of China (Grants No. U25A6008(3)) and National Key Research and Development Program of China (Grants No. 2022YFA1404503 and No. 2021YFC2201802), and the Central Government Guidance Funds for Local Science and Technology Development Projects (Grant No. YDZJSX2025D001).
\bibliography{apssamp} 
\end{document}